\documentclass[%
reprint,
amsmath,
amssymb,
aps,prl
]{revtex4-1}

\usepackage{graphicx}
\usepackage{dcolumn}
\usepackage{bm}
\usepackage{amsmath}
\usepackage{xcolor}
\usepackage{hyperref}
\usepackage{hhline}

\usepackage{diagbox}
\usepackage{siunitx}
\usepackage{enumitem}

\usepackage[normalem]{ulem}
\newcommand\redout{\bgroup\markoverwith
	{\textcolor{red}{\rule[.5ex]{2pt}{1pt}}}\ULon}

\newcommand{\rmnum}[1]{\uppercase\expandafter{\romannumeral #1}}

\begin{document}
	
\title{Weak-field Hall Resistivity and Spin/Valley Flavor Symmetry Breaking in MAtBG}

\author{Ming Xie}
\author{A. H. MacDonald}
\affiliation{Physics Department, University of Texas at Austin, Austin TX 78712}

\date{\today}

\begin{abstract}
	Near a magic twist angle, the lowest energy conduction and valence bands of bilayer graphene 
	moir\'e superlattices become extremely narrow.  The band dispersion that remains is sensitive to the moir\'e's strain pattern, nonlocal tunneling between layers, and filling-factor dependent Hartree and exchange band renormalizations.
	In this article we analyze the influence of these band-structure details on the pattern of flavor-symmetry-breaking observed in this narrow band system, and on the associated pattern of Fermi surface reconstructions revealed by weak-field-Hall
	and Schubnikov-de Haas magneto-transport measurements. 
\end{abstract}

\maketitle

\textit{Introduction.}---\noindent
When twisted close to a magic \cite{BMModel} orientation angle, the ground state of bilayer graphene
exhibits \cite{CaoInsulator,CaoSuper,YoungDean,Efetov} a rich series of strongly correlated electronic ground states. The magic-angle twisted-bilayer graphene (MAtBG) phase diagram is most strongly dependent on
twist angle $\theta$ and on moir\'e band filling factor $\nu = n A_M$, where $n$ is the carrier density and $A_M$ is the area of the moir\'e pattern unit cell, but is also responsive to other external parameters
including the orientation angles of the encapsulating hexagonal boron nitride layers, 
and the vertical separation between the bilayer and the electrical gate or gates used to manipulate 
the carrier density.  

\begin{figure}[t]
	\includegraphics[width=0.99\linewidth]{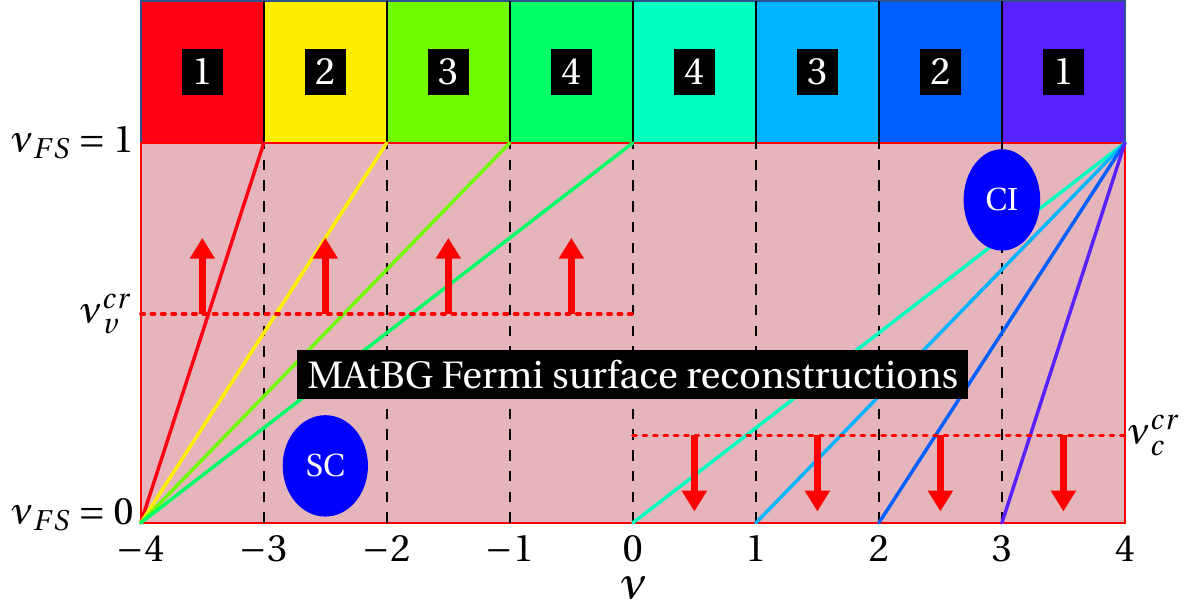}
	\caption{\label{schematic} 
		The flat bands of MAtBG are partially occupied for  filling factor $\nu \in (-4,4)$ 
		and exhibit flavor symmetry breaking over much of this range.  
		While not entirely universal, flavor symmetry breaking tends to follow the following rules. For negative carrier density
		flavor symmetry breaking depopulates flat valence bands so as to keep the partial filling factors
       $\nu_{\small FS}$ of the remaining partially occupied flavors, which have a Fermi surface, above $\nu^{cr}_{v} \sim 0.55$ as indicated by the red arrows.  
		For positive carrier densities flavor symmetry breaking favors complete occupation of one, two, or three 
		conduction bands so as to keep  $\nu_{\small FS}$ of the remaining partially occupied flavors
		below $\nu^{cr}_{c} \sim 0.2$.  The solid lines plot the band filling factor per partially occupied flavor in $1$, $2$,
		$3$, and $4$ Fermi surface states.  This pattern of flavor symmetry breaking allows for insulating states at all non-zero $\nu$ 
		between -4 and 4, and for Chen insulators (CI) at odd integer $\nu$.
		The strongest superconductivity (SC) seems to emerge from states with two valence band
		Fermi surfaces, and the strongest anomalous Hall effects seem to occur in states with one conduction band 
		Fermi surface.}
		\end{figure}

Experimental work over the past couple of years
\cite{Efetov,GoldhaberGordon,YoungQAHE,Andrei, Pasupathy, Yazdani1, Nadj-Perge, Yazdani2, Ilani, Petr, YoungOrbital} has established that the spin/valley flavor symmetries responsible for the four-fold degeneracy of the moir\'e bands are often broken when the flat conduction band is partially occupied 
or the flat valence band is partially emptied. 
The flavor symmetry breaking is reminiscent of the 
behavior of Bernal-stacked bilayer graphene in a strong magnetic field when its flat $N=0$ Landau levels are
partially filled \cite{McCann2006, Barlas2008,Cote2010, Jeil2011, Barlas2012, Feldman2009, Martin2010, Weitz2010}.
The pattern of symmetry breaking is however quite distinct in the two cases.  Instead of filling up the eight bands one 
at a time to minimize the exchange energy, as observed in the quantum Hall case,
the flavor symmetry breaking in MAtBG has a different guiding principle, which is illustrated schematically in Fig.~\ref{schematic}.
The broken symmetry states in MAtBG almost always conspire to keep the partial filling factors $\nu_{FS}$ of all flavors with a valence band Fermi surface
 above a critical value $\nu^{cr}_{v}$ typically $\sim 0.55$, 
and those of all flavors with a conduction band Fermi surface below $\nu^{cr}_{c} \sim 0.2$.
The observations from which this general rule is abstracted are briefly surveyed in the supplemental material~\cite{supplemental}.
Because the maximum conduction band electron Fermi surface areas are smaller than the maximum valence band 
hole Fermi surfaces areas, flavor symmetry breakings are more abundant on the conduction band side than on the valence band side.
In this Letter we provide an explanation for this behavior that is based on the influence of 
strain, nonlocal tunneling between layers, and filling-factor dependent Hartree and exchange interactions on the 
band structure.  Our analysis is informed by Schubnikov-de Haas and weak-field Hall magneto-transport data 
\cite{CaoInsulator, YoungDean, GoldhaberGordon, Efetov, YoungQAHE, AndreiHall,Petr,Nadj-Perge2020, Zeldov, EfetovChernInsulator, YoungCI, YoungHofstadter, YoungIsospin}.

\textit{MAtBG Bandstructure.}---\noindent
Magic-angle strong correlation physics 
persists over a small range of twist angles, covering perhaps $\sim 0.2^{\circ}$, within which
the typical velocity within the flat bands is reduced by an order of magnitude or more relative to the band velocity in an isolated graphene layer.  We argue here, however, that the band dispersion that survives near the magic twist angle plays a crucial role in establishing the ground state phase diagram.

We specify the single-particle MAtBG flat bands using four phenomenological 
parameters, $w_{AB}$ - the interlayer intersublattice tunneling strength, $w_{AA}$ - the interlayer intrasublattice tunneling strength, a tunneling non-locality parameter $w_{NL}$, and $\epsilon^{-1}$, a parameter 
that characterizes how strongly the moir\'e band Hamiltonian
is modified by interactions.  
The spinless single-particle Hamiltonian projected onto valley K takes the form:
\begin{align}
	\mathcal{\hat{H}}^{\rm{K}}_{sp} = 
	\begin{pmatrix}
		h_{\theta/2}(\bm{k}) & T(\bm{r}, \bm{r}')\\
		T^\dagger(\bm{r}, \bm{r}')  & h_{-\theta/2}(\bm{k}')
	\end{pmatrix}
	\label{nonlocalHam}
\end{align}
where $\hat{h}_{\pm \theta/2}$ are the Dirac Hamiltonians for isolated rotated graphene top (+) and bottom (-) layers
and $T(\bm{r}, \bm{r}')$ is the inter-layer tunneling term.
The ratio $w_{AA}/w_{AB}$ has the value $1$ for simplified models \cite{BMModel} 
in which the graphene layers are rotated rigidly, but is known \cite{Carr2018, Carr, ChiralModel}
to be reduced relative to $1$ when strain and corrugation are taken into account. 
The non-locality of the interlayer tunneling is known \cite{Carr, Fang2019, Walet2019} to be principally responsible 
for particle-hole asymmetry in MAtBG, which we model by setting 
\begin{align}
	T(\bm{r}, \bm{r}') = \sum_{j=0}^2 \sum_{\bm{p}}T_j(\bm{p})  e^{-i\bm{q}_j\cdot (\bm{r}+\bm{r}')/2} e^{i\bm{p}\cdot(\bm{r}-\bm{r}')} 
\end{align}
where $T_j(\bm{p})= t(\bm{p}) T^{BM}_j$ and the momentum dependent tunneling amplitude 
$t(\bm{p}) = t_{k_D} + (dt/dp)_{p=k_D}(k_D-p)$.
We parametrize the slope of the momentum dependence in $t(\bm{p})$ by $w_{NL}\equiv(dt/dp)_{p=k_D}|\bm{b}|$.
We note that $T(\bm{r}, \bm{r}')=T(\bm{r})\delta(\bm{r}, \bm{r}')$ is local in the Bistritzer-MacDonald (BM) model
where the tunneling matrix $T^{BM}_j=\omega_{AA}/\omega_{AB}+ \cos(j\phi)\sigma_x + \sin(j\phi)\sigma_y$
is independent of momentum (see Supplemental Material~\cite{supplemental}).
Below we explain what magneto-transport measurements tell us about the typical values of these band parameters.

\begin{figure}[t]
	\includegraphics[width=0.95\linewidth]{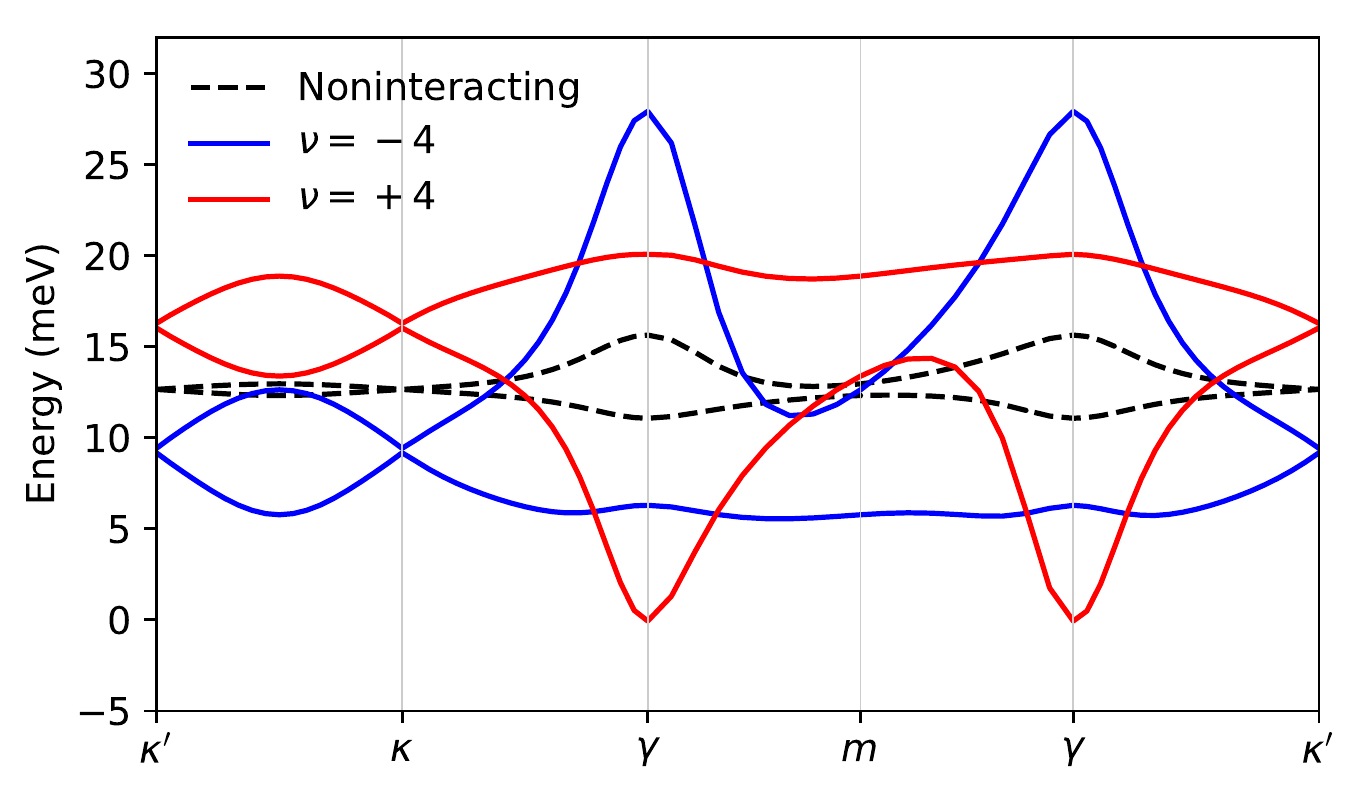}
	\caption{\label{fullandempty} Band-structures in the full and empty moir\'e band limits for band parameters 
		$\hbar v k_{\theta} = 1.69 w_{AB}$, corresponding to a twist angle of about $\theta\sim1.1^{\circ}$.  
		$w_{AA}/w_{AB} = 0.6$, $w_{NL}=20$ meV and $\epsilon^{-1}=0.03$.  Interaction effects strongly weaken
		dispersion when the Fermi level moves toward $\gamma$ on either electron or hole sides.}
\end{figure}

The flat bands of MAtBG have a simple and systematic dependence on band filling, one that 
we argue plays an important role in the flat-band phase diagram. In Fig.~\ref{fullandempty} 
we plot the flat bands when they are completely full and when they are empty.
These calculations neglect mixing between flat and remote bands, an approximation that is 
justified by flat-band spectral isolation.  In this approximation the many-electron ground state is fully determined by the Pauli exclusion principle in both limits, and is a single Slater determinant in which single-particle-state occupations numbers are either $0$ or $1$.  It follows that the electron self-energy is given exactly by the Hartree-Fock operator:
\begin{align}
	\hat{\Sigma}^{HF} = \hat{\Sigma}^{H}(\delta\rho) + \hat{\Sigma}^{F}(\delta\rho),
\end{align}
where $\delta\rho=\sum'_{\alpha, n, \bm{k}} |\alpha, n, \bm{k}\rangle \langle\alpha, n, \bm{k}|-\rho_{iso}$ 
is the ground state density matrix defined relative to neutral isolated graphene states,
$\alpha$ is a composite label for valley and spin, $n$ is a single-particle band label,
the prime restricts the summation to filled bands, and 
$\hat{\Sigma}^{H}$ and $\hat{\Sigma}^{F}$
are the usual Hartree and Fock self-energies (See Supplemental Material~\cite{supplemental} for further detail.)
In Fig.~\ref{fullandempty} we have used $\epsilon^{-1}=0.03$ to account \cite{GWscreening} for screening by the surrounding 
dielectric, by the two-dimensional material itself, and by the gates.

The bands plotted in Fig.~\ref{fullandempty} are eigenvalues of the band Hamiltonian
$\hat{\mathcal{H}}_{\cal B}=\hat{\mathcal{H}}_{sp} + \hat{\varSigma}^{HF}$, where $\hat{\cal H}_{sp}$ is the single-particle moir\'e band Hamiltonian.  
The difference between the quasiparticle dispersions in the empty and full band limits is due to the differences between the self-energy operator when the flat bands are full, $\hat{\Sigma}^{HF}_f$, and when the flat bands are empty, $\hat{\Sigma}^{HF}_e$.  
We see in Fig.~\ref{fullandempty} that the effect of the self-energy is to
lower (raise) the energy near the center of the moir\'e Brillouin zone ($\bm{k}=\bm{\gamma}$), 
relative to those near the corners of the Brillouin zone ($\bm{k}=\bm{\kappa},\bm{\kappa'}$) 
as the flat conduction (valence) bands are filled (emptied). 
This behavior has been explained \cite{ourHF, Guinea} in terms of Hartree interactions and the concentration of flat band states at AA positions in the bilayer moir\'e pattern, where the wavefunctions of flat band states near $\bm{\gamma}$ have lower weight \cite{Po2019, Song2019}.  
Exchange interactions also play a role in reshaping the bands, and a more critical role in breaking flavor symmetries (see \cite{ourHF} and Supplemental Material~\cite{supplemental}).
These interaction effects have a smooth dependence on filling factor which justifies the approximation
\begin{equation}
	\hat{\cal H}_{\cal B} =\hat{\cal H}_{sp}+ \frac{1}{2}\big[ \hat{\Sigma}^{HF}_f +  \hat{\Sigma}^{HF}_e + \frac{\nu}{4}(\hat{\Sigma}^{HF}_f - \hat{\Sigma}^{HF}_e)\big].
\end{equation}
The end result is that Fermi velocities are extremely small when the bands are nearly empty and nearly full, in contrast to the case of single-particle band models for which Fermi velocities are maximized for nearly full and nearly empty bands.
This property is captured only when self-energies from frozen remote bands are included in the theory.  
We argue below that it also plays a crucial role in determining the pattern of flavor symmetry breaking.

\textit{Schubnikov-de Haas oscillations.}---\noindent
Oscillations in physical properties associated with periodic filling and emptying of 
nascent Landau levels in weak magnetic fields have long \cite{Schoenberg} been used to measure the Fermi surfaces of metals.
In two-dimensional materials the most accessible oscillations are normally those of longitudinal resistance referred to as Schubnikov-de Haas (SdH) oscillations.  
SdH oscillations in MAtBG
\cite{CaoInsulator, YoungDean, GoldhaberGordon, Efetov,YoungQAHE,AndreiHall,Petr,Nadj-Perge2020,Zeldov,EfetovChernInsulator,YoungCI, YoungHofstadter, YoungIsospin, Petr2020}
are generally speaking observable only near neutrality for 
$\nu \in (-1.6,~0.8)$ and for $|\nu| \in (2,3)$.  We attribute this property to the interaction induced reductions in Fermi velocity when the bands are nearly empty or nearly full as explained above.  
Comparison of Fermi surface area to carrier density suggests that all four flavors have
equivalent Fermi surfaces near neutrality.  For $|\nu| \in (2,3)$ 
the same comparison suggests that only two of the four flavors have Fermi
surfaces, the other two flat bands having apparently been depopulated on the valence 
band side and completely filled on the conduction band side. 
These observations are consistent with the interpretation of weak-field Hall observations discussed below, which are able to provide valuable information over the full range of flat-band filling factors because they do not rely on an adequate Landau level spacing.

\textit{Weak-field Hall resistivity.}---\noindent
When mean-free-paths exceed Fermi wavelengths, the transport properties of two-dimensional 
Fermi liquids can be described using Boltzmann transport theory.  Employing a relaxation time approximation, a practical necessity when the source of scattering is unknown,
the conductivity tensor of a system with $C_6$ symmetry is given  \cite{Ong} to leading order in magnetic field $B$ by
\begin{align}
	\sigma_{xx}  &= e^2\tau  \sum_{\bm{k}}\left(-\frac{\partial f_{\bm{k}}}{\partial E}\right)  v_{x}^2, \notag\\
	\sigma_{yx}  &= 
	\frac{2e^3\tau^2 B}{\hbar} \sum_{\bm{k}} \left(-\frac{\partial f_{\bm{k}}}{\partial E}\right) 
	v_x (\bm{v}\times \bm{z}) \cdot \bm{\nabla}v_y,     \notag\\
	\sigma_{xx}&=\sigma_{yy},\hspace{1em} \sigma_{yx} = - \sigma_{xy}
\end{align}
It follows that to first order in $B$ the Hall resistivity
\begin{equation}
	\label{eq:rhoxy}
	\rho_{xy} = \frac{\sigma_{yx}}{\sigma_{xx}^2} 
	\equiv \frac{B}{n_H ec}.
\end{equation}
In Eq.~\ref{eq:rhoxy} the coefficient of the $B$-linear term in $\rho_{xy}$ is 
a pure bandstructure property because the scattering time $\tau$ cancels between numerator and 
denominator.  As suggested by the final form on the right-hand-side, it is convenient to 
characterize this quantity by the Hall density, $n_H$, which is defined by this equation.
$|n_H|$ equals the carrier density for isotropic Fermi surfaces \cite{Ong}.

\begin{figure}[t]
	\includegraphics[width=\linewidth]{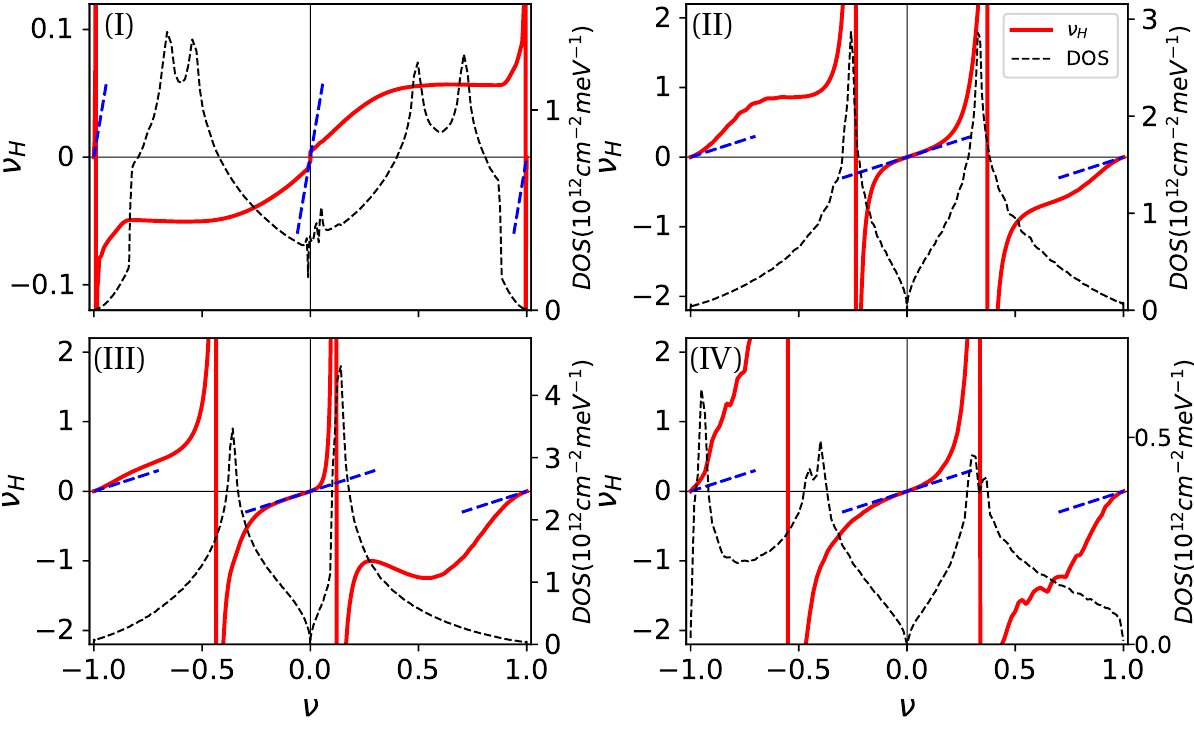}
	\caption{
		Hall filling factor $\nu_H = n_H A_M$ {\it vs.} band filling factor $\nu$ of a single flavor for the band models
		(\rmnum{1}-\rmnum{4}) defined in the main text. 
		In numerical order the four models improve the BM model (\rmnum{1}) by sequentially adding corrections 
		for strain (\rmnum{2}), nonlocal tunneling (\rmnum{3}), and interactions (\rmnum{4}).
       The black dashed curves plot the band DOSs, whose peaks are located at the VHS energies.
		The blue dashed lines are the Hall densities $\nu_H=\nu+1, \nu, \nu-1$ for isotropic Fermi surfaces 
		near $v=-1,0,1$ respectively.}
	\label{Fig:HallDensity}
\end{figure}

In Fig.~\ref{Fig:HallDensity} we plot Hall filling factor $\nu_H=n_HA_M$ {\it vs.} filling factor $\nu$
for partially occupied MAtBG bands at $w_{AB}/\hbar v K = 1.69$ for four different bandstructure models: 
(\rmnum{1})$w_{AA}=w_{AB},~w_{NL}=0$, 
(\rmnum{2})$w_{AA}/w_{AB}=0.6,~w_{NL}=0$,
(\rmnum{3})$w_{AA}/w_{AB}=0.6,~w_{NL}=20$ meV, and 
(\rmnum{4})$w_{AA}/w_{AB}=0.6,~w_{NL}=20$ meV with $\epsilon^{-1}=0.03$.
The first model is the  BM model specified in~\cite{BMModel},
which is improved in models (\rmnum{2}), (\rmnum{3}), and (\rmnum{4}) by sequentially 
adding corrections for strain and corrugation, nonlocal tunneling, and interactions.
In all cases $\nu_H \sim \nu$ close to neutrality, expected for isotropic Fermi surfaces.
Away from neutrality, Fermi surfaces are more anisotropic for larger $\omega_{AA}/\omega_{AB}$ 
and this leads to larger deviations \cite{Ong} of the Hall density from the corresponding Fermi surface area curves marked in blue.
(The sensitivity of the Hall density to band parameters is discussed in detail in the Supplemental Material~\cite{supplemental}.) 

The most prominent features in Fig.~\ref{Fig:HallDensity} are the switches 
between large $\nu_H$s of opposite signs that occur once for positive and once for negative filling factors. The filling factors at which these sign changes occur are close to the filling factors at which 
the topology of the Fermi surfaces changes from electron-like to hole-like. 
As explained in the Supplemental Material~\cite{supplemental}, the sign changes do not precisely 
match the van Hove singularities (VHS) at which the flat band 
density of states (DOS) diverges logarithmically yielding a feature that is prominent
in tunneling spectroscopy measurements \cite{Andrei, Yazdani1, Pasupathy, Yazdani2, Nadj-Perge}. 
We see in Fig.~\ref{Fig:HallDensity} that the positions of these 
sign changes are sensitive to the bandstructure model details.
They move very close to neutrality when the strain/corrugation corrections are added.
Including nonlocal tunneling shifts the position closer to (further from) neutrality on the conduction (valence) band side, 
strongly violating particle-hole symmetry.
Interaction renormalizations move the sign change positions on both sides away from neutrality 
by an amount determined by the strength of interaction.

\textit{Hall Density, Fermi Surface Reconstruction, and Liftshitz Transitions.}---\noindent
The Hall densities in Fig.~\ref{Fig:HallDensity} differ qualitatively from experimental data,
which typically exhibit five or more jumps in value for $\nu \in (-4,4)$ compared to the two jumps present in Fig.~\ref{Fig:HallDensity} 
assuming no flavor symmetry breaking.
Some of these jumps are evidently due to spin/valley flavor symmetry breaking phase transitions, which reconstruct the Fermi surfaces. 
These transitions sequentially maximize hole 
densities for one, two, or three flavors on the hole side, and electron densities 
of one, two, or three flavors on the conduction band side.
The guiding principle for these phase transitions seems to be to place the Fermi levels of the partially occupied 
bands, whenever possible, on the side of the van Hove singularity closer to the Dirac point of that flavor.
We can understand this behavior \cite{metamagnetism} qualitatively as a combined consequence of the exchange energy gained by flavor polarization 
and the favorable band energy per particle when the Fermi level is placed in the low-density-of-states region near the 
Dirac point.

\begin{figure}[t]
	\centering
	\includegraphics[width=\linewidth]{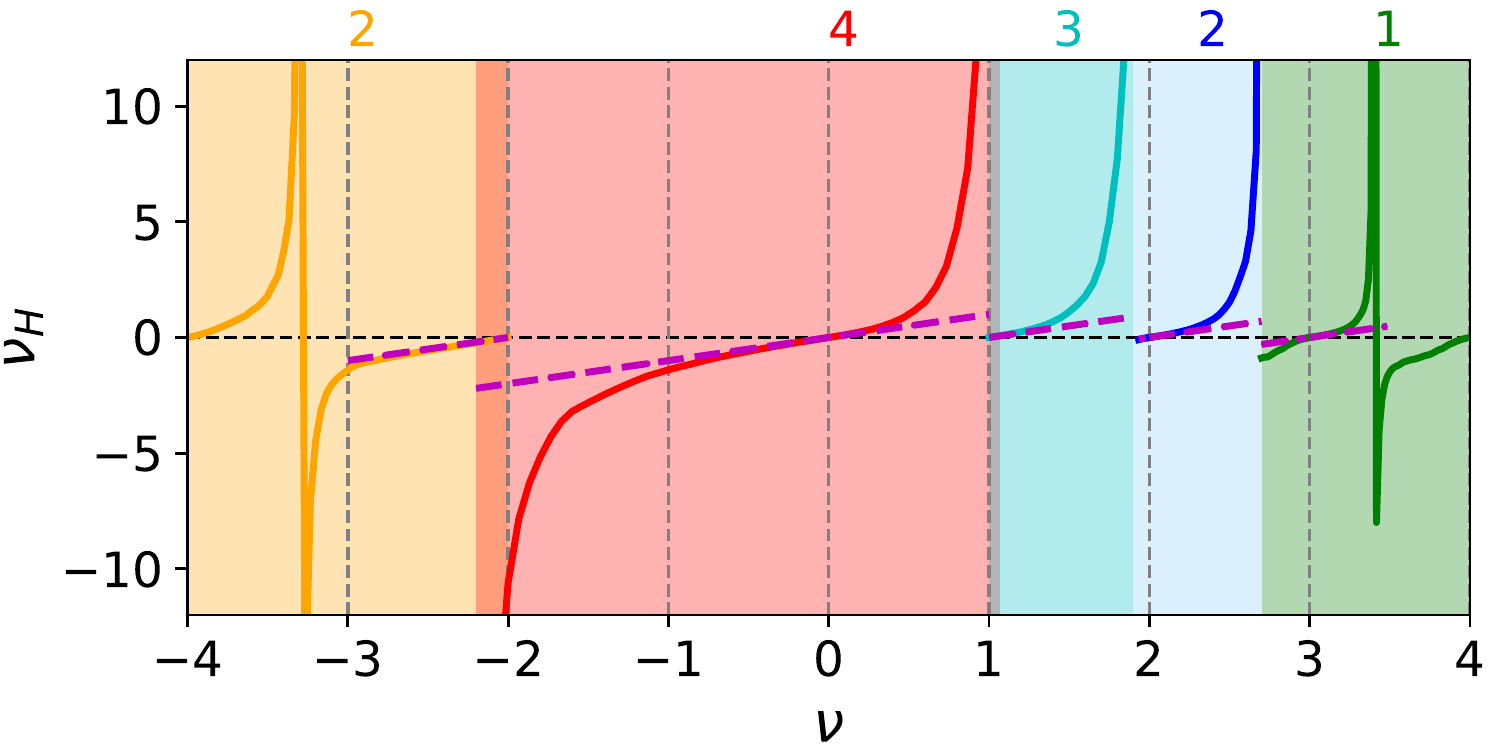}
	\caption{Hall filling factor as a function of band filling factor allowing spin/valley flavor
	symmetry breaking, calculated using interaction renormalized bands and the pattern of 
	flavor symmetry explained in the main text. The shading specifies the number of partially occupied bands,
	{\it i.e.} the number of Fermi surfaces in each region of filling factor.  The solid lines mark the 
	Hall density that would be calculated if the Fermi surfaces were approximated as isotropic.
	}
	\label{Fig:TheoryHallCurve}
\end{figure}

The number of additional Hall density jumps and their precise filling-factor positions are somewhat 
sample dependent.
Fig.~\ref{Fig:TheoryHallCurve} shows calculated Hall density that accounts for flavor symmetry breaking transitions. 
The second Hall density jump on the valence band side is consistent with traversal of VHSs in doubly degenerate Fermi surfaces.  
On the conduction band side the first Hall density jump occurs already near $\nu\approx 4\nu^{cr}_c$.  
We attribute the difference between electrons and holes to the difference in the position at which the VHS occurs.  
Two additional jumps in Hall density typically occur on the conduction band side, 
and each seems to be associated with a flavor depopulation event.  
One important consequence is that states with an odd number of Fermi surfaces are more common on the conduction band 
side. 
We have constructed the theory curve in Fig.~\ref{Fig:TheoryHallCurve} from
Fig.~\ref{Fig:HallDensity} by summing Hall conductivity contributions over flavors.  As shown there, we are 
able to explain the experimental results by infering that 
there is one flavor symmetry breaking transition on the hole side at 
$\nu_{FS}\approx\nu^{cr}_v$ (or equivalently $\nu\approx-4(1-\nu^{cr}_v)$)
from a four-Fermi surface state to a two-Fermi surface state, 
and that there are three flavor-symmetry-breaking 
transitions on the electron side, at $\nu\approx 4\nu^{cr}_c, 1+3\nu^{cr}_c, 2+2\nu^{cr}_c$,
to three, two, and one Fermi surface states.
Because the Hall density curves are qualitatively dependent on the pattern of flavor symmetry breaking,
as illustrated in the Supplementary material, this interpretation can be made with 
considerable confidence.

The flavor symmetry breaking behavior evident in the weak-field Hall and SdH data 
is understandable in terms of mean-field considerations.  States that have a Fermi level on the neutrality side of the 
VHS are favored by a low DOS close to the Fermi level \cite{farsideVHS}.  Particle-hole asymmetry can be 
explained by the closer proximity of VHS to neutrality on the conduction band side, which forces flavor symmetry 
breaking earlier in the conduction band filling process than in the valence band emptying process, and therefore 
favors states with an odd number of Fermi surfaces that can host topologically non-trivial states 
\cite{GoldhaberGordon, YoungQAHE, Yazdani2020Chern}.

\textit{Discussion.}---\noindent
This Letter addresses the implications of weak-field magneto-transport data for the correlation physics of MAtBG.
The moir\'e filling factor ranges over which SdH oscillations are visible identify where Fermi level quasiparticles have the largest velocities.  The fact that SdH is most visible near neutrality is at first sight surprising since the independent electron flat bands are most dispersive in precisely the opposite limit, namely for nearly full or nearly empty bands.  We have interpreted this behavior as a consequence of interaction-induced band renormalizations that flatten the conduction band top and the valence band bottom 
when the Fermi level is near these band edges.

Weak-field Hall effect measurements provide additional information since this quantity is observable
at larger quasiparticle masses and stronger disorder.  We have concluded that the many jumps in Hall density
that typically occur as a function of band filling cannot be explained on the basis of single-particle physics,
since these allow for only one jump each side of the neutrality point.
We attribute the additional jumps to a series of flavor-symmetry breaking phase transitions that reconstruct Fermi 
surfaces by redistributing the occupancy among flavors.  These reconstructions favor partially occupied bands 
whose Fermi surfaces are on the neutrality side
of the VHS present in each band. 
A similar conclusion was reached on the basis of thermodynamic compressibility measurements in Ref~\cite{Ilani}.  Our analysis argues that filling-factor dependent band renormalizations play an essential role in the robustness of this effect, and that nonlocal interlayer tunneling controls its particle-hole asymmetry which is substantial in most cases.  

Unlike a ferromagnetic metal, in which the spin-dependence of electron-electron interactions is normally ignored, the valley and orbital dependence of interactions in MAtBG could \cite{ValleyWave, Dai2019} play as important a role as the band Hamiltonian in the anisotropy energy scales, and could play a role in the superconducting state that emerges at the
lowest temperatures.  Surveying the experimental literature and assigning Fermi surface degeneracies via considerations like those discussed in this MS, it seems that superconductivity occurs with similar transition temperatures in states with two and four Fermi surfaces, but has much lower transition temperatures when seen in states with an odd number of Fermi surfaces.

\begin{acknowledgments}
	{\em Acknowledgment.}---\noindent	The authors acknowledge helpful interactions with Eva Andrei, Dmitry Efetov, 
	Wei Qin, Petr Stepanov, Shuang Wu, Andrea Young, and Zhenyuan Zhang. 
	We thank Stephen Carr for a helpful conversation explaining the relationship between 
	particle-hole asymmetry in the flat bands and tunneling non-locality.  
	This work was supported by the U.S. Department of Energy, Office of Science, Basic Energy Sciences, under Award No. DE-SC0022106.
\end{acknowledgments}

\renewcommand\thefigure{S\arabic{figure}}
\setcounter{figure}{0}
\renewcommand\theequation{S\arabic{equation}}
\setcounter{equation}{0}

\newpage
\appendix
\widetext
\renewcommand\thefigure{S\arabic{figure}}
\setcounter{figure}{0}
\renewcommand\theequation{S\arabic{equation}}
\setcounter{equation}{0}
\newcommand\titlelowercase[1]{\texorpdfstring{\lowercase{#1}}{#1}}

\newcommand\ins{INS}
\newcommand\ci{INS}
\newcommand\cs{SM}
\newcommand\ds{SM}
\newcommand\norm{\hspace{0.2em}-----\hspace{0.2em}}
\newcommand\res{RES}
\newcommand\rev{REV}
\newcommand\lin{LC}
\newcommand\abs{Absent}
\newcommand\hcc{\vspace{1em}}

\newcommand\EfetovI{Ref.[5]}     
\newcommand\EfetovII{Ref.[14]}   
\newcommand\EfetovIII{Ref.[27]}  
\newcommand\EfetovIV{Ref.[16]}  
\newcommand\YoungI{Ref.[4]}     
\newcommand\YoungII{Ref.[7]}     
\newcommand\YoungIII{Ref.[29]}  
\newcommand\YoungIV{Ref.[30]}  
\newcommand\YoungV{Ref.[31]}  
\newcommand\JiaI{Ref.[32]} 
\newcommand\PabloI{Ref.[2]}      
\newcommand\AndreiI{Ref.[25]}  

\newpage
\begin{center}
	{\bf Supplemental Material for \\ ``Weak-field Hall Resistivity and Spin/Valley Flavor Symmetry Breaking in  MAtBG''}
\end{center}

\section{Summary of weak-field-Hall and Schubnikov-de Haas measurements }

Signatures of spin/valley flavor symmetry breaking 
have been routinely seen in magic angle twisted bilayer graphene
since the breakthrough discovery of correlated insulating states and superconductivity a few years ago [2].
Correlated insulating states are manifestations of broken flavor symmetries at integer filling factors,
and are observable directly via transport experiments.
Away from integer filling factors,  flavor symmetry breaking transitions result in Fermi surface reconstructions,
which can be detected by more sensitive experimental probes, such as
weak-field Hall and Schubnikov-de Haas measurements as discussed in the main text.
Despite frequent observations of these signatures, however, 
variations from sample to sample in where and how many transitions occur 
across the full filling factor range $\nu\in(-4,4)$ persist.
As we argued in the main text, details of the flat band dispersion, which are influenced by strain, nonlocal tunneling and interaction renormalizations, 
play a crucial role in determining the pattern of flavor symmetry breaking.
In addition there are extrinsic factors such as gate screening and disorder that may hinder the
observation of flavor symmetry breaking and are difficult to control precisely at present.
In this supplemental section, by surveying past experimental works which have relatively more complete weak-field Hall and Schubnikov-de Haas data,
we summarize the overall trend in the pattern of flavor symmetry breaking,
in an effort to disambiguating the extrinsic effects mentioned above from 
the bandstructure effects discussed in this Letter.

Table I lists observed features in both longitudinal resistance and weak-field Hall density measurements at integer moir\'e band filling factors.
We categorize experimental features in resistance measurements into `insulating' (INS) and `semimetallic' (SM)
(see the table caption for their precise definitions),
and denote cases with none or poorly developed features by a long dash (---).
In addition, we do not distinguish between trivial and Chern insulators and label both as INS.
Weak-field Hall density measurement often show rich features across the full filling factor range.
Here we recognize three main features: `reset' (RES), `reversal' (REV), and `linear crossing' (LC).
The latter two are typical features when the chemical potential passes through a van Hove singularity and a Dirac point, respectively.
Similar to the Hall density case, we label absence of features or insufficiently developed features by a long dash (---).

\begin{table}[t!]
	\centering
	\caption{\textbf{Survey of low-temperature transport and weak-field Hall experiments at integer filling factors of MAtBG.}
		Observed features in longitudinal resistivity and weak-field Hall density measurements are included as the first and the second term in each table cell respectively.
		Features in resistivity measurements are categorized into either `insulating' (INS), when the resistivity shows a thermally activated behavior,
		or semimetallic (SM), when the resistivity peaks at an integer but exhibits metallic temperature dependence.
		Features in Hall density measurements are categorized into reversal (REV), when the Hall density jumps from a positive to a negative value,
		reset (RES), when Hall density reset to zero, and linear crossing (LC), when Hall density crosses zero continuously and linearly as $\nu$ 
		is varied. A long dash (------) means these is no or only insufficiently developed feature. 
		A table cell with only one term means the corresponding Hall density data is not available. $d_{\rm gate}$ is distance to nearby metallic gates.
	}\par
	\medskip
	\begin{tabular}{{|>{\centering\arraybackslash}m{0.08\textwidth}}*{7}{|>{\centering\arraybackslash}m{0.097\textwidth}}|c|c|}
		\hhline{|=|=|=|=|=|=|=|=|=|=|}
		\diagbox[height=2.9\line, width=4.85em]{\hspace{0.6em}$\theta$}{$\nu$\hspace{0.5em}} 
		&            $-3$       &        $-2$          &       $-1$              &       $0$     &    $1$                 &             $2$         &        $3$             & $d_{\rm gate}$& Reference            \\ 	\hhline{|=|=|=|=|=|=|=|=|=|=|}
		\hcc 1.08$^\circ$   &   \norm                &   \ins               &   \norm                   & \cs             & \norm                & \ins                       &  \norm               &                 &\PabloI-D1                    \\ \hline           
		\hcc 1.16$^\circ$    & \cs,\norm           &   \ins, \res      &   \norm,\norm       & \cs, \lin     & \norm,\norm      & \ins, \res              &  \cs,\norm        &                 &\PabloI-D4                    \\ \hline 
		\hcc 1.17$^\circ$    &   \cs,\rev             &   \ins, \res      &   \norm,\norm       & \ds, \lin      &  \norm,\norm    &  \ins, \res             &  \cs,\rev            &                & \AndreiI                  \\ \hline  
		\hcc 1.10$^\circ$     &  \cs, \norm        &  \ins, \res          &  \cs, \norm          &  \ins, \lin   &  \ins, \norm      &  \ins, \res            &  \ins, \norm       &12.5nm   &\EfetovI-D1$^{a)}$ \\ \hline
		\hcc 1.15$^\circ$     &   \norm, \rev      &  \cs, \res          &  \norm, \norm    &  \ins, \lin    &  \norm, \norm  &  \norm, \norm   &  \norm, \rev       &7nm         &\EfetovII-D1            \\ \hline
		\hcc 1.04$^\circ$    &   \norm                  &  \norm             &   \norm                 &  \ins          &   \norm             &  \norm                     &  \norm            &9.8nm     & \EfetovII-D2           \\ \hline
		\hcc 1.04$^{\circ~b)}$&   \norm, \rev &   \norm, \res  &   \norm, \norm   &  \cs,  \lin     & \norm,\norm    & \norm, \res        & \norm, \rev       &                & \EfetovIII-A1           \\ \hline
		\hcc 1.03$^\circ$    &   \norm             &   \ins                 &   \norm                & \cs              & \cs                      &  \ins                   & \cs                      &               & \EfetovIII-A2          \\ \hline
		\hcc 1.08$^\circ$    &   \norm, \rev      &   \ins, \res        &   \norm,\norm     & \ins, \lin     & \ci, \res              &  \ins, \res          & \ins, \res             & 7nm       & \EfetovIV                  \\ \hline
		\hcc 1.14$^\circ$     &   \norm, \rev       &   \ins,\norm    &   \norm,\norm    & \ins, \lin     & \norm                &  \ins,\norm         & \ins,\rev            &$>$30nm  & \YoungI-D1             \\ \hline          
		\hcc 1.27$^{\circ~c)}$& \norm, \rev      &   \ins, \res        &   \norm,\norm    & \cs, \lin     & \cs, \res             &  \ins, \res           & \ins,\norm         &$>$30nm& \YoungI-D2             \\ \hline          
		\hcc 1.15$^\circ$     &        \norm           &   \norm            &  \norm                 & \ins            & \norm                &  \cs                       & \ci                     & $>$40nm   & \YoungII                  \\ \hline       
		\hcc 1.08$^\circ$    &  \norm, \norm   &   \ins, \res       &   \norm,\norm     & \cs, \lin     & \ins, \res            &  \ins, \res            & \ins,\norm         &68nm    & \YoungIII-D1          \\ \hline 
		\hcc 1.09$^\circ$   &         \norm          &   \ins                &   \norm                 & \cs             & \norm                &  \ins                    & \ins                      &6.7nm    & \YoungIII-D2         \\ \hline          
		\hcc 1.04$^\circ$   &  \norm,\norm     &\norm,\norm   &   \norm,\norm      & \cs, \lin     & \cs,\norm          &  \ins, \res           & \ins,\norm          &38nm     & \YoungIII-D3         \\ \hline 
		\hcc 1.18$^\circ$    &   \norm, \rev       &\norm,\norm   &   \norm,\norm      & \cs, \lin     & \norm,\norm    &  \norm,\norm    & \norm,\norm       &7.5nm    & \YoungIII-D4         \\ \hline 
		\hcc 1.12$^\circ$    &    \cs, \res           &   \ins, \res       &   \norm,\norm       & \cs, \lin    & \norm,\norm     &  \ins, \res            & \ins, \norm         &45nm     & \YoungIII-D5$^{d)}$ \\ \hline 
		\hcc 1.06$^\circ$   & \norm,\norm      &   \ins, \res      &   \norm,\norm       & \cs, \lin     & \cs, \norm         &  \ins                      &  \ins                    &30nm & \YoungV-D2$^{e)}$  \\ \hline
		\multicolumn{10}{l}{\vphantom{$\frac{A}{A}$}a). See also \EfetovII-D3.}\\
		\multicolumn{10}{>{\centering\arraybackslash}l}{b). The weak-field Hall data on this sample is taken at a field of $B=1T$, larger than the value of $B=0.5T$ used by most of}\\
		\multicolumn{10}{>{\centering\arraybackslash}l}{\hphantom{b). }the experiments listed here. See text for discussions.}\\
		\multicolumn{10}{>{\centering\arraybackslash}l}{c). This sample is brought to magic angle condition  by hydrostatic pressure.}\\
		\multicolumn{10}{>{\centering\arraybackslash}l}{d). See also \YoungIV and \YoungV-D1.}\\
		\multicolumn{10}{>{\centering\arraybackslash}l}{e). See also \JiaI.}\\
	\end{tabular}
\end{table}


We focus on flavor symmetry breaking transitions at zero temperature and zero magnetic field. 
Although the base temperatures in these experiments are in the 10 mK-100 mK range,
we assume no transition will occur upon further lowering the temperature and the results can be extrapolated to zero temperature.
The weak-field Hall experiments listed here were mostly done under a small perpendicular magnetic field on the order of $0.5T$.
Because magnetic field is known to enhance flavor symmetry breaking through coupling to spin and orbital magnetizations,
caution needs to be taken when interpret these experiments as a measure of zero-field ground state properties, 
and one should also examine the magnetic field dependence.
On the other hand, SdH experiments reveal properties of the metallic states at non-integer filling factors, 
such as the number of flavors or Fermi surfaces which cross the Fermi energy,
which show directly the reconstruction of the Fermi surface associated with flavor symmetry breaking.
However, one need also to keep in mind that quantum oscillations in MAtBG develop typically at 
at fields $B>0.5T$ and transitions observed there are usually finite field properties.
%
%

Combining transport, weak-field Hall and SdH oscillation data,
we observe following trends in flavor symmetry breaking:
\begin{itemize}
	\item Insulating states are more abundant on the conduction band side ($\nu>0$) than on the valence band side ($\nu<0$).
	Insulating states at both $\nu=2$ and $\nu=-2$ are more stably observed than those at $|\nu|=1,3$,
	and insulating states are more often seen at $\nu=1,3$ than at $\nu=-1,-3$.
	\item Flavor symmetry breaking transitions often occur at non-integer filling factors and the resulting broken symmetry persists across next integer
	(as $|\nu|$ is increased from 0) leading to insulating states. 
	
	\item On the valence band side, upon hole doping, the first flavor symmetry breaking transition occurs after $\nu=-1$ and slightly before $\nu=-2$.
	No insulating state has been observed so far at $\nu=-1$ although quantum oscillations can emerge under strong enough magnetic field. 
	Near  $\nu=-3$,  a `reversal' (REV) in hall density due to passing through a VHS is been observed dominantly. 
	In samples exhibiting very clean quantum oscillations, a reset in Hall density tends to occur at $\nu=-3$.
	\item On the conduction band side, upon increasing $\nu$, the first flavor symmetry breaking transition occurs before $\nu=1$, much closer to neutrality.
	The $\nu=3$ insulator is more frequently observed compared to $\nu=1$ insulator.
	
	\item Quantum oscillations are stably seen emerging from $\nu=0,\pm2$ and appear dominantly on the side of these integers that is further away from neutrality.
	Quantum oscillations emerging from $\nu=\pm1,\pm3$ are usually span lesser range of the $(B,\nu)$ space and are seen less frequently.
\end{itemize}
%

In the ideal case when the quasiparticle gap of half filled flavors is fully exposed 
and the quasiparticle bands of fully filled or fully empty flavors are shifted away by exchange interaction,
one expects an `insulating' behavior to concur with a Hall density `reset'.
However, experiments sometimes show a `insulating' state without a Hall density `reset', or a Hall density `reset' accompanied instead by a semimetallic state.
These can be viewed as a ``weaker" version of the flavor symmetry breaking insulator where
quasiparticle gap of half-filled flavors is not fully exposed but instead masked, on the $\nu<0$ side, by the quasiparticle bands of nearly empty flavors,
and on the $\nu>0$ side, by the quasiparticle bands of nearly full flavors.
In this sense, they are semimetals. One should also keep in mind that other possibilities, such as disorder and formation of flavor domains,
can also be responsible.

\section{Effective nonlocal tunneling model}
We generalize the continuum BM model to account for the nonlocal nature of interlayer tunneling in twisted bilayer graphene, 
which is shown by {\it ab initio} calculations to be responsible for the particle-hole asymmetry.
In the original BM model, the interlayer tunneling term is taken to be momentum independent which means
that in real space electrons can only tunnel vertically between the same in-plane locations of the top and bottom layers.
In other words, the interlayer tunneling term between electrons at $\bm{r}$ in the top layer and at $\bm{r}'$ in the bottom layer
is diagonal in a coordinate representation: $T(\bm{r}, \bm{r}')=T(\bm{r})\delta(\bm{r}, \bm{r}')$.
We generalize this tunneling model to allow tunneling which is off-diagonal in in-plane coordinates,
\emph{i.e.} tunneling that is nonlocal.  
The spinless single-particle Hamiltonian projected onto a single valley takes the form: 
\begin{align}
	\mathcal{\hat{H}}^{\rm{K}}_{sp} = 
	\begin{pmatrix}
		h_{\theta/2}(\bm{k}) & T(\bm{r}, \bm{r}')\\
		T^\dagger(\bm{r}, \bm{r}')  & h_{-\theta/2}(\bm{k}')
	\end{pmatrix}
	\label{nonlocalHam}
\end{align}
where $\hat{h}_{\pm \theta/2}$ are the Dirac Hamiltonians for isolated rotated graphene top (+) and bottom (-) layers,
\begin{align}
	h_{\theta}(\bm{k}) = -\hbar v_D |\bar{\bm{k}}| 
	\begin{pmatrix}
		0 & e^{i (\theta_{\bar{\bm{k}}}- \theta)} \\
		e^{-i  (\theta_{\bar{\bm{k}}}- \theta)}  & 0
	\end{pmatrix},
\end{align}
$\theta_{\bar{\bm{k}}}$ is the orientation angle of momentum measured from the Dirac point of corresponding layer
$\bar{\bm{k}}=\bm{k}-\bm{K}_{\pm\theta/2}$.  $\bm{K}_{\pm\theta/2}$ is the Dirac momentum of top/bottom layer.
We choose the convention in which $\bm{k}$ and $\bm{k}'$ are measured from a common origin of momentum.

The interlayer tunneling Hamiltonian takes the form
\begin{align}
	T(\bm{r}, \bm{r}') = \sum_{j=0}^2 \sum_{\bm{p}}T_j(\bm{p})  e^{-i\bm{q}_j\cdot (\bm{r}+\bm{r}')/2} e^{i\bm{p}\cdot(\bm{r}-\bm{r}')}.
	\label{off-diagT}
\end{align}
The average coordinate experiences the moir\'{e} periodic modulation which is accompanied by momentum boosts 
$\bm{q}_j=\bm{0}$, $\bm{b}_1$ or $\bm{b}_2$ for $j=0, 1, 2$ respectively.
$\bm{b}_{1,2}=(\pm 1/2,\sqrt{3}/2)4\pi/(\sqrt{3}a_M)$
are the basis vectors of moir\'{e} reciprocal lattice, where $a_M=a/(2\sin(\theta/2))$ is the lattice constant of moire pattern 
and $a$ the lattice constant of monolayer graphene.
The dependence on the relative coordinate is captured by the momentum dependent tunneling matrices:
\begin{align}
	T_j(\bm{p})&= t(\bm{p}) T^{BM}_j, \\
	t(\bm{p}) &= t_{k_D} - \frac{dt}{dp}|_{p=k_D}(p-k_D),\\
	T^{BM}_j&=\frac{\omega_{AA}}{\omega_{AB}} \sigma_0 +  \left[\cos(j\phi)\sigma_x + \sin(j\phi)\sigma_y\right].
\end{align}
In the BM model, $t(\bm{p}) \equiv t_{k_D}$.
Here we retain momentum dependence in $t(\bm{p})$ to first order in $\bm{p}$ and assume that it is 
independent of the orientation of $\bm{p}$.
We parameterize the first derivative using $\omega_{NL} \equiv dt/dp \cdot|\bm{b}|$ where
$|\bm{b}|$ is the length of basis vectors in moir\'e reciprocal space.
The average magnitude of interlayer tunneling is taken to be $\omega_{AB}\equiv t_{k_D}=110$ meV
and $\omega_{AA}/\omega_{AB}=0.6$ if not otherwise specified.

It is important to note that $\bm{p}$ is momentum measured from the $\bm{\Gamma}$ point
and implicitly depends on the direction of momentum boost or the index $j$ as detailed below.
In the extended zone scheme, interlayer tunneling preserves momentum $\bm{p}=\bm{k}+\bm{G}_j=\bm{k}'+\bm{G}'_j$ given that 
$\bm{k}$ and $\bm{k}'$ are also measured from $\bm{\Gamma}$.
$\bm{G}_j$ and $\bm{G}'_j$ are reciprocal lattice vectors of the rotated top and bottom graphene layers which satisfy
$\bm{G}_0=\bm{G}'_0=0$, $\bm{G}'_2-\bm{G}_2=\bm{b}_2$ and $\bm{G}'_1-\bm{G}_1=\bm{b}_1$.
Because we are interested in a small region with a size on the order of mBZ in momentum space near the Dirac points, 
$|\bm{k}-\bm{K}_{\pm\theta/2}|\sim |\bm{b}|/|\bm{K}_{\pm\theta/2}|\sim\theta$,
the relevant reciprocal vectors $\bm{G}=m\bm{G}_1+n\bm{G}_2$ are only $\bm{G}_0$, $\bm{G}_1$, and $\bm{G}_2$.

The value of $\omega_{NL}$ can be determined experimentally, for example, by measuring the single-particle energy gap 
between the flat bands and the remote bands. As we detail in the section below, the difference between the 
gaps to remote conduction bands and remote valence bands is directly related to $\omega_{NL}$.
We take $\omega_{NL}$ to be $\sim 20$meV, which approximately produces the correct single-particle gap sizes.
In the next section we discuss the effect of nonlocal tunneling on the energy bandstructure.

\section{Symmetry of the nonlocal tunneling Hamiltonian and particle-hole asymmetry}
The original single-particle continuum model has an approximate particle-hole symmetry
that becomes exact when we ignore the phase factor $e^{\pm i\theta/2}$ due to the rotation of top and bottom layers'
momentum space or when the interlayer intra-sublattice tunneling amplitude set to zero ($\omega_{AA}=0$),
as in chiral limit models. In reality, $\omega_{AA}$ is neither not equal to $0$, or even close to $0$.
Instead, \textit{ab initio} calculations estimate that $\omega_{AA}/\omega_{AB}$ is in the range $0.6-0.7$.
Therefore we consider the more general and realistic case with nonzero $\omega_{AA}$.
Because the symmetry of the intralayer Dirac Hamiltonian remains intact, we focus on the symmetry of the 
interlayer tunneling Hamiltonian when it is generalized to become nonlocal.

The real space representation $\mathcal{D}$ of a symmetry transformation $\hat{\mathcal{S}}$ involving space group element $\hat{g}$ in general depends on spatial coordinates, $\mathcal{D}_{\hat{S}}\equiv\mathcal{D}_{\hat{S}}(\bm{r})$.
For nonlocal operators, the transformation works as  
\begin{align}
	\mathcal{O}'(\bm{r}, \bm{r}') = \mathcal{D}_{\hat{\mathcal{S}}}(\bm{r}) 
	\mathcal{O}(\hat{g}\bm{r}, \hat{g}\bm{r}') \mathcal{D}^{-1}_{\hat{\mathcal{S}}}(\bm{r}')
\end{align}

The nonlocal tunneling Hamiltonian in Eq.~\ref{nonlocalHam} preserves the original model's $\hat{\mathcal{C}}_2\hat{\mathcal{T}}$ symmetry,
where $\hat{\mathcal{C}}_2$ is two-fold rotation around $z$ axis and $\hat{\mathcal{T}}$ is the spinless time-reversal transformation:
\begin{align}
	\mathcal{D}_{\hat{\mathcal{C}}_2\hat{\mathcal{T}}} T(\hat{\mathcal{C}}_2\bm{r}, \hat{\mathcal{C}}_2\bm{r}') \mathcal{D}^{-1}_{\hat{\mathcal{C}}_2\hat{\mathcal{T}}}
	= \hat{\sigma}_x T^*(-\bm{r}, -\bm{r}') \hat{\sigma}_x
	= T(\bm{r}, \bm{r}'), \label{C2Tsymm}
\end{align}
where $\mathcal{D}_{\hat{\mathcal{C}}_2\hat{\mathcal{T}}}=\sigma_x\mathcal{K}$ is independent of $\bm{r}$ and layer,
and $\mathcal{K}$ is the complex conjugation operator.
The last equality can be obtained using Eq.~\ref{off-diagT} keeping in mind that $T^{BM}_j$ is invariant under $\mathcal{D}_{\hat{\mathcal{C}}_2\hat{\mathcal{T}}}$.

Three-fold rotational symmetry $\hat{\mathcal{C}}_3$ (around $z$ axis) is also retained because
\begin{align}
	\mathcal{D}^t_{\hat{\mathcal{C}}_3} (\bm{r})T(\hat{\mathcal{C}}_3\bm{r}, \hat{\mathcal{C}}_3\bm{r}') (\mathcal{D}_{\hat{\mathcal{C}}_3}^{b}(\bm{r}'))^{-1}
	= T(\bm{r}, \bm{r}').
	\label{C3}
\end{align}
where $	\mathcal{D}^t_{\hat{\mathcal{C}}_3}(\bm{r})=e^{i\frac{2\pi}{3}\sigma_z} e^{i\bm{q}_1\cdot\bm{r}}$
for the top layer and 
$	\mathcal{D}^b_{\hat{\mathcal{C}}_3}(\bm{r}')=e^{i\frac{2\pi}{3}\sigma_z} e^{i(\bm{q}_1-\bm{q}_2)\cdot\bm{r}}$ for the bottom layer.
The tunneling amplitude $t(\hat{\mathcal{C}}_3 \bm{p})=t(\bm{p})$ is invariant under $\hat{\mathcal{C}}_3$.

In addition, the nonlocal tunneling Hamiltonian also preserves the two-fold rotational symmetry around the $x$ axis,
\begin{align}
	\mathcal{D}_{\hat{\mathcal{M}}_x}
	\begin{pmatrix}
		0 & T(\bm{r}, \bm{r}') \\
		T^\dagger(\bm{r}, \bm{r}') & 0
	\end{pmatrix}
	\mathcal{D}_{\hat{\mathcal{M}}_x}^{-1}
	=
	\begin{pmatrix}
		0 & T(\bm{r}, \bm{r}') \\
		T^\dagger(\bm{r}, \bm{r}') & 0
	\end{pmatrix}.
\end{align}
where $\hat{\mathcal{M}}_x$ changes $(x, y)$ to $(x,-y)$ and swaps both sublattice and layer as $\mathcal{D}_{\hat{\mathcal{M}}_y }=\sigma_x\tau_x$.
$\tau$ is Pauli matrix for layer pseudo-spin.
In deriving this equality, we have used $\hat{\mathcal{M}}_x \bm{q}_{1,2}$=$-\bm{q}_{2,1}$.

Finally we come to the original model's particle-hole symmetry under the transformation 
$\hat{\mathcal{P}}$ that applies when the phase factors $e^{i\pm\theta/2}$ are ignored.
$\hat{\mathcal{P}}$ operates on the tunneling Hamiltonian as
\begin{align}
	\mathcal{D}^t_{\hat{\mathcal{P}} }(\bm{r})
	T(\hat{\mathcal{M}}_x\bm{r}) 
	(\mathcal{D}^b_{\hat{\mathcal{P}} }(\bm{r}))^{-1}
	= - 
	T(\bm{r})
	\label{ph}
\end{align}
where $\mathcal{D}^{t/b}_{\hat{\mathcal{P}} }(\bm{r})=\pm\sigma_xe^{2i(\bm{q}_1-\bm{q}_2)\cdot\bm{r}}$
and $\hat{\mathcal{M}}_y$ changes $(x,y)$ to $(-x,y)$.
Eq.~\ref{ph} follows directly from the relation $\hat{\mathcal{M}}_y\bm{q}_{0}=\bm{q}_{0}$ and $\hat{\mathcal{M}}_y\bm{q}_{1,2}=\bm{q}_{2,1}$.
When generalized to include a nonlocal contribution, the tunneling Hamiltonian is no longer particle-hole symmetric and 
instead satisfies
\begin{align}
	\mathcal{D}^t_{\hat{\mathcal{P}} }(\bm{r})
	T(\hat{\mathcal{M}}_x\bm{r}, \hat{\mathcal{M}}_x\bm{r}') 
	(\mathcal{D}^b_{\hat{\mathcal{P}} }(\bm{r}'))^{-1}
	= - 
	T(\bm{r}, \bm{r}') e^{2i(\bm{q}_1-\bm{q}_2)\cdot(\bm{r}-\bm{r}')}.
\end{align}
In summary, the nonlocal tunneling Hamiltonian preserves the original model's $\hat{\mathcal{C}}_2\hat{\mathcal{T}}$ symmetry, 
three-fold rotation symmetry and mirror symmetry. However, it violates the particle-hole symmetry.
In the next section, we discuss quantitative changes in the single-particle energy dispersion due to 
tunneling non-locality.

\section{Bandstructure of the nonlocal tunneling model}

\begin{figure}[t]
	\includegraphics[width=0.99\linewidth]{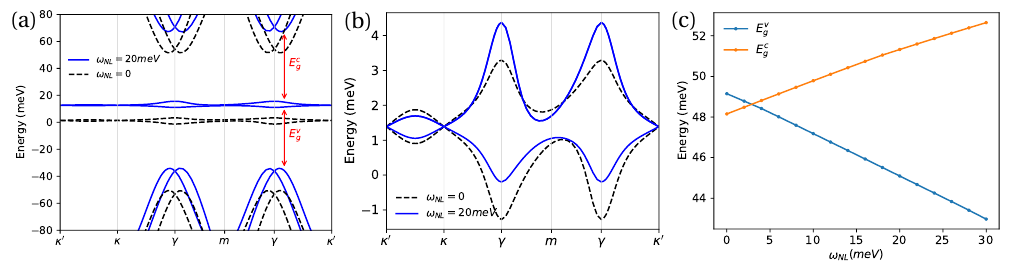}
	\caption{Bandstructure of the nonlocal tunneling model. 
		(a) Comparison of bandstructures between local $\omega_{NL}=0$ (black dashed) and nonlocal $\omega_{NL}=20meV$(blue solid) models. 
		$\theta=1.1$ and $\omega_{AA}/\omega_{AB}=0.6$ are used in both cases.
		The red arrows mark the single-particle gaps $E^v_g$ and $E^c_g$. (b) Flat-band zoom for the bandstructure in (a). 
		For comparison purposes, the blue solid curve is shifted in energy so that the two cases match in energy at the Dirac points.	
		(c) The single-particle gaps $E^v_g$ and $E^c_g$ as a function of the strength of nonlocality $\omega_{NL}$.
		The small $e^{\pm i\theta/2}$ phase factor is included in both cases.
	}\label{fig:dispersion}
\end{figure}

In this section, we show how the nonlocal tunneling modifies the single-particle bandstructure.
Fig.~\ref{fig:dispersion}(a) plots the typical bandstructure with $\omega_{NL}=20$meV.
The bands shift upward in energy relative to those of the original model, which is approximately symmetric around zero energy
with a small asymmetry due to the $e^{\pm i\theta/2}$ factor.
The single particle gap is reduced on the valence band side but increased
on the conduction band side.
The flatband dispersion changes as the valence band becomes flatter while the conduction band becomes wider.
The detailed change in flatband dispersion is also manifested in its Fermi surface topology as discussed in next section.
As the nonlocality parameter $\omega_{NL}$ is increased, the single particle gap to the remote bands 
on the conduction/valence band side
increases/decreases almost linearly, as shown in Fig.~\ref{fig:dispersion}(c).
The reversal of the relative size of $E^c_g$ and $E^v_g$ indicates that $e^{\pm i\theta/2}$ factor causes 
particle-hole asymmetry of the opposite sense compared to the larger nonlocal tunneling.


\section{Fermi surface topology and weak-field Hall resistivity}

\begin{figure}[t!]
	\includegraphics[width=0.75\linewidth]{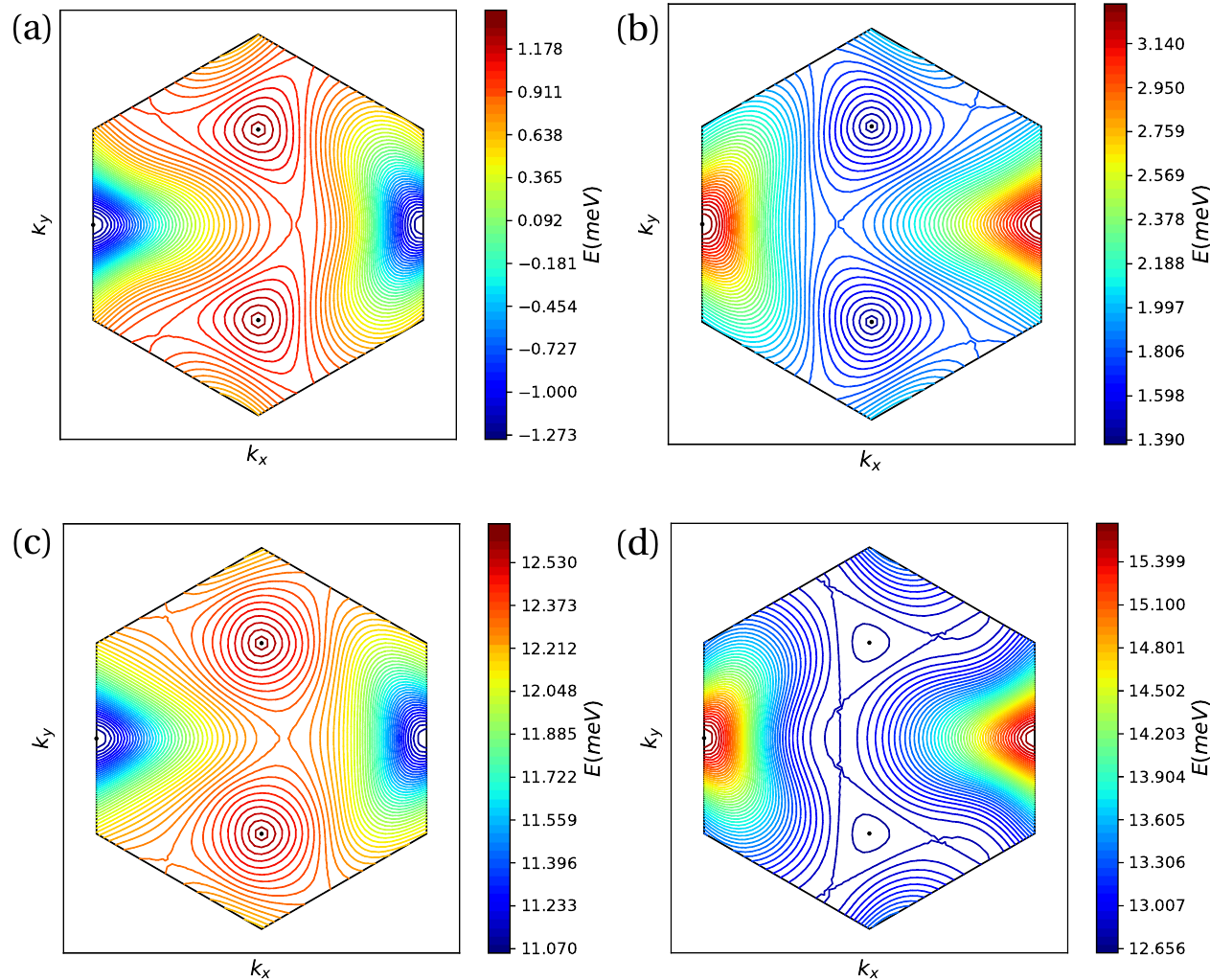}
	\caption{Fermi surface contours of single-particle models (a, b)  without $\omega_{NL}=0$ and (c, d) with nonlocal tunneling term $\omega_{NL}=20$meV .
		(a, c) are for flat valence bands and (b, d) for flat conduction bands.
		Same parameters are used as in Fig.~\ref{fig:dispersion}(a, b).
		The color of each contour line represents its Fermi energy and the energy difference between neighboring contours is constant
		so the density of contour lines indicates the steepness of the dispersion.
	}
	\label{FScontour}
\end{figure}

The topology of the flat-band Fermi surface is directly tied to many experimental observations 
including SdH oscillations 
and magneto-transport measurements.
It is also crucial for understanding  the pairing mechanism of superconductivity.
Up to the present, knowledge of the flat-band Fermi surface is still very limited. 
In this section, we start by studying how nonlocal tunneling influences the Fermi surface topology.
We ignore interaction effects for the moment so the energy dispersion is independent of filling factor.

Fig.~\ref{FScontour} plots Fermi surface contours, i.e equal-energy contours of the single-particle continuum 
model with and without nonlocal tunneling corrections.
The flat-band Fermi surface without nonlocal tunneling (Fig.~\ref{FScontour} (a,b))
is characterized by a single electron (hole) pocket centered around the $\bm{\gamma}$ point in momentum space,
when the Fermi energy is near the bottom (top) of the valence (conduction) band.
As the size of the Fermi surface increases, its shape  changes gradually from circular-like to triangular-like.
On the other hand, when the Fermi energy is slightly below (above) the Dirac energy,
the Fermi surface has two hole (electron) pockets centered around the two Dirac points
and their shapes also changes from circular-like to triangular-like.
The two sides meet at a saddle-point van Hove singularity (VHS) point,
at which $\gamma$ centered triangle-like Fermi pockets
and $\kappa,\kappa'$ centered Fermi pockets touch..

With nonlocal tunneling (Fig.~\ref{FScontour} (c,d)), the symmetry between conduction and valence flat bands is noticeably broken.
Although the valence band Fermi surface contours do not change much, those of the 
conduction flat band change qualitatively.
At the conduction band bottom, in addition to electron pockets near the two Dirac points, 
additional pockets emerge near $m$ as a result of the distortion of $\gamma$ pockets which reaches 
the mBZ boundary before becoming connected with the Dirac pockets.

\begin{figure}[b!]
	\includegraphics[width=\linewidth]{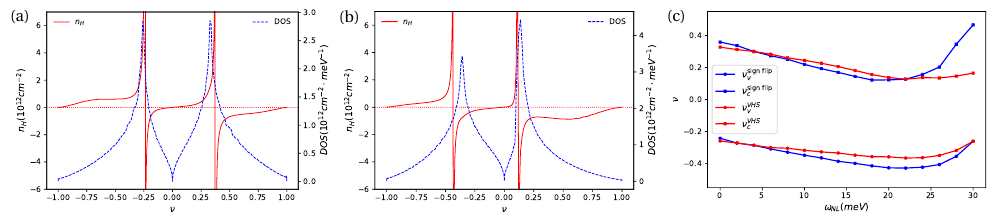}
	\caption{Weak-field Hall density (red solid) and density of states (blue dashed) of single-particle flat bands 
		without (a) and with (b) nonlocal tunneling {it vs.} band filling factor $\nu$.
		The parameters are the same as in Fig.~\ref{fig:dispersion}(a).
		(c) Locations of Hall density sign changes and of VHSs as a function of nonlocal tunneling 
		parameter $\omega_{NL}$.}	
	\label{nHDOS}
\end{figure}

As mentioned in the main text, the filling factor position at which Hall density changes sign does not coincide exactly with the VHS.
Fig.~\ref{nHDOS}(a,b) shows two typical examples of Hall density and DOS of the single particle band without and with nonlocal tunneling as the filling factor is varied.
The sign change position and the VHS point deviates slightly from each other.
Fig.~\ref{nHDOS}(c) plots the sign change and VHS positions on both valence and conduction band sides as a function of the nonlocal tunneling parameter.

\section{Hartree and Fock self-energy}
The Hartree and Fock self-energies are calculated using unscreened 
and $SU(4)$ flavor symmetric Coulomb interactions,
$V_{\alpha\beta}(\bm{q})= 2 \pi e^2/(\epsilon |\bm{q}|) \exp(-qd_{\alpha_l,\beta_l})$,
where $\alpha,\beta$ are composite indices of layer ($l$) and sublattice ($s$).
$\epsilon$ is the effective dielectric constant which we take as a control parameter 
for interaction strength.
$d_{\alpha_l,\beta_l}=d(1-\delta_{\alpha_l,\beta_l})$ accounts for the vertical distance
between $\alpha$ and $\beta$ sublattices with layer index $\alpha_l$ and $\beta_l$ respectively. $d$ is the interlayer spacing.
In layer and sublattice basis, the Hartree and Fock self-energy operators take the forms
\begin{align}
	\Sigma^{H,s,\mu}_{\alpha, \bm{G}; \beta, \bm{G}'}(\bm{k}) 
	&= \frac{1}{A} \sum_{\alpha'}
	V_{\alpha'\alpha}(\bm{G}'-\bm{G}) 
	\delta \rho_{\alpha'\alpha'}(\bm{G}-\bm{G}')  \delta_{\alpha\beta},\\
	\Sigma^{F, s,\mu}_{\alpha, \bm{G}; \beta, \bm{G}'}(\bm{k})
	&=-\frac{1}{A}\sum_{\bm{G}'', \bm{k}'}
	V_{\alpha\beta}(\bm{G}''+\bm{k}'-\bm{k}) 
	\delta \rho^{s,\mu}_{\alpha, \bm{G}+\bm{G}'';\beta, \bm{G}'+\bm{G}''} (\bm{k}').
\end{align}
Here $\bm{k}$ is understood to be restricted
to the first moir\'{e} Brillouin zone (mBZ) ($\bm{k} \in$ mBZ).
$\bm{G}=m\bm{b}_1+n\bm{b}_2$ is moir\'e reciprocal vectors with integers $m,n$.
$s$ and $\mu$ are labels of spin and valley.
$\delta\rho \equiv \rho  -\rho_{\rm{iso}}$ is the density matrix defined relative to that
of isolated rotated graphene layers each filled up to the charge neutrality point.
Because we do not consider coherence between different valleys or spins, $\delta\rho=\sum_{s,\mu}\delta\rho^{s,\mu}$.
As in the main text, $\delta\rho_e$ is the density matrix of filling all the remote valence bands and 
keeping the flat bands empty. $\delta\rho_f$ is the density matrix of filling all the remote valence and flat bands.

\begin{figure}[t!]
	\includegraphics[width=0.99\linewidth]{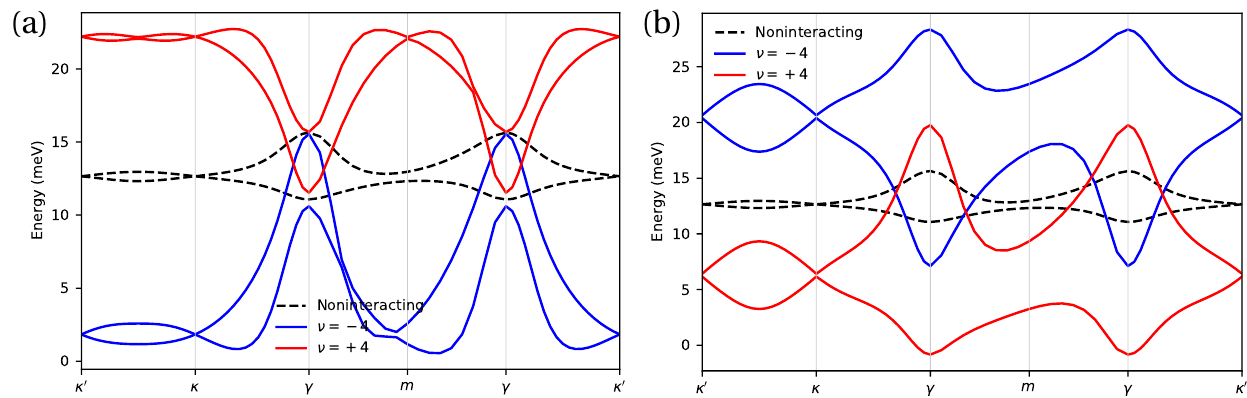}
	\caption{Effects of Hartree (a) and Fock (b) self-energy renormalization on quasiparticle dispersions. The parameters used are same as in Fig. 2 of main text. See Fig. 1 for the total effect of Hartree and Fock terms.}
	\label{H-F-effect}
\end{figure}

We then project the self-energy onto the flatband Hilbert space defined by the eigenvectors,
\begin{align}
	|\psi_{s, \mu, n,\bm{k}}\rangle= \sum_{\alpha} z^{n, s,\mu}_{\alpha, \bm{G},\bm{k}} |\alpha,\bm{G},\bm{k}\rangle,
\end{align}
where $\pm$ represents valley $K/K'$ and $n=c,v$ for conduction and valence bands.
The self-energy operator projected onto the flatband Hilbert is therefore 
\begin{align}
	\hat{\Sigma}^{H}(\delta\rho) = \sum_{s,\mu, n, n', \bm{k}} \langle \psi_{s, \mu, n,\bm{k}}|\Sigma^{H,s,\mu}(\bm{k})|\psi_{s, \mu, n',\bm{k}}\rangle
	\hat{c}^\dagger_{s,\mu, n, \bm{k}}\hat{c}_{s,\mu, n', \bm{k}}, \\
	\hat{\Sigma}^{F}(\delta\rho) = \sum_{s,\mu, n, n', \bm{k}} \langle \psi_{s, \mu, n,\bm{k}}|\Sigma^{F,s,\mu}(\bm{k})|\psi_{s, \mu, n',\bm{k}}\rangle
	\hat{c}^\dagger_{s,\mu, n, \bm{k}}\hat{c}_{s,\mu, n', \bm{k}}.
\end{align}
The Hartree self-energy of hole (electron) doped state lowers (raises) the quasiparticle energy at $\kappa,\kappa'$ relative to that at $\gamma$.
As shown in Fig.~\ref{H-F-effect}(a), the $\gamma$ point energy is kept unchanged because the electron wavefunction has vanishing weight on AA position in real space
where the Hartree self-energy attracts (repels) the most.
The Fock self-energy also renormalizes quasiparticle bands by increasing the band width.
More importantly, it shifts the whole bands up when doped with holes and down when doped with electrons. (Fig.~\ref{H-F-effect}(b)). 
The total energy can be lowered when doping bands in a flavor-dependent manner, which is responsible for
spin and/or valley flavor symmetry breaking.

\begin{figure}[t!]
	\includegraphics[width=0.9\linewidth]{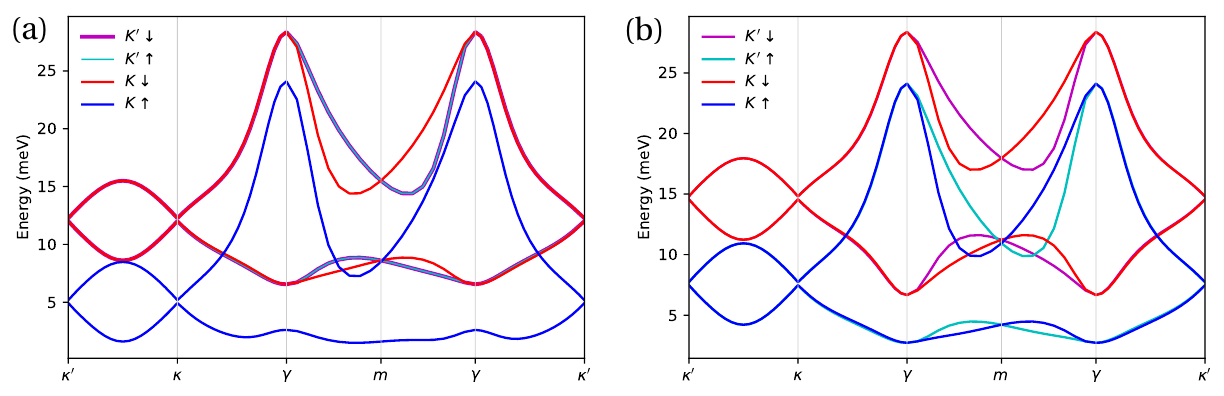}
	\caption{ Quasiparticle bands  of (a) $\nu=-1$ state with both spin and valley polarization and (b) $\nu=-2$ ferromagnetic state.  
		The nonlocal parameter is $\omega_{NL}=20$meV.
		The pattern of broken symmetries and Hall density jumps can be explained by retaining strain, non-locality, 
		and interaction band modifications. }
	\label{num2bands}
\end{figure}

Because the interaction self-energy is filling-factor dependent, the quasiparticle dispersion changes
as the flat band is filled or emptied. 
In Fig.~\ref{num2bands}, we plot the quasiparticle dispersions of the ferromagnetic state at $\nu=-2$ and of the
spin-valley polarized state at $\nu=-3$, which are both single-Slater determinant states.

\begin{figure}[b!]
	\includegraphics[width=0.7\linewidth]{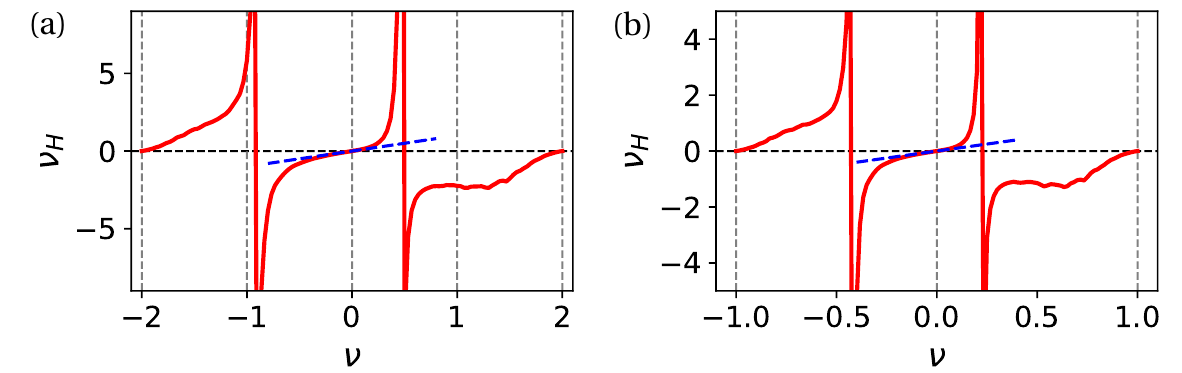}
	\caption{Hall filling factor $\nu_H$ as a function of filling factor $\nu$ for flavor symmetry breaking states with two (a) and one (b) flavor occupied. 
		(a) is assuming flavor symmetry breaking states with ferromagnetic order and (b) with both spin and valley polarization. We only show the range of 
		filling factors near neutrality. The parameters used here are $\omega_{AA}/\omega_{AB}=0.6$, $\omega_{NL}=20$meV and $\epsilon^{-1}=0.01$.}
	\label{Hall_SM}
\end{figure}

Finally, we show examples of Hall density as a function of filling factor for flavor symmetry breaking states that has only two or one Fermi surfaces.
The Hall density is calculated for filling factor dependent quasiparticle bandstructures at their corresponding Fermi levels.
From Fig.~\ref{Hall_SM} we see that the overall behavior is similar to that of the flavor symmetric case shown in Fig.3(c) with a slight shift of 
VHS mainly due to the difference in Hartree self-energy which includes contribution from all occupied flavors.

\begin{thebibliography}{99}
\bibitem{farsideVHS}
The even lower DOS interval that occurs for nearly full and nearly empty bands when interactions are neglected 
does not share the enhanced stability of its Dirac point cousin because of interaction renormalizations that reduce 
Fermi velocities when bands are nearly full or nearly empty.
\end{thebibliography}
\end{document}